\documentclass{article}
\usepackage{amsmath}
\usepackage{ipe}

\newcommand{\aroba}{\symbol{64} }

\newenvironment{algorithm}[1]{\noindent\ \\
  \mbox{{\bf Algorithm {#1}}} \\
	[-0.1in]}{\vspace{-0.1in}\mbox{{\bf End of the Algorithm}}\smallskip}

\newtheorem{theorem}{Theorem}
\newtheorem{thlemma}[theorem]{Lemma}
\newtheorem{fact}[theorem]{Fact}
\newtheorem{thcorollary}[theorem]{Corollary}

\newcommand{\proof}[1]{ {\bf Proof.} #1 \hfill$\diamondsuit$}

\begin{document}
\title{COMPUTING  LARGEST CIRCLES SEPARATING TWO SETS OF SEGMENTS\thanks{
	Research of J. Czyzowicz and J. Urrutia are supported by NSERC.
	Short version of this paper appeared in 8CCCG, 1996.
                                }}
\author{
JEAN-DANIEL BOISSONNAT\thanks{Jean-Daniel.Boissonnat{\aroba}sophia.inria.fr}
 \and
JUREK CZYZOWICZ\footnote{czyzowicz{\aroba}uqah.uquebec.ca}
\and
OLIVIER DEVILLERS\footnote{Olivier.Devillers{\aroba}sophia.inria.fr}
\and
JORGE URRUTIA
\and
MARIETTE YVINEC\footnote{Mariette.Yvinec{\aroba}sophia.inria.fr}
}
\maketitle

\begin{abstract}
A circle $C$ separates two planar sets if it encloses one of the sets and
its open interior disk does not meet the other set. A separating circle is
a largest one if it cannot be locally increased while still separating
the two given sets. An $\Theta(n \log  n)$ optimal algorithm is proposed to
find all largest
circles separating two given sets of line segments when line segments are allowed
to meet only at their endpoints.  In the general case, when line
segments may intersect $\Omega(n^2)$ times, our algorithm can be adapted
to work
in $O(n \alpha(n) \log n)$ time and $O(n \alpha(n))$ space,
where $\alpha(n)$ represents the extremely slowly growing inverse of the
Ackermann function.
\end{abstract}

\section{Introduction}

Let $\cal C$ denote a family of Jordan curves in the plane. 
Two sets $P$ and $Q$ in the plane are $\cal C$-separable if there exists
$\xi \in \cal C$, such that every point of one of these sets lies in the 
closed region inside $\xi$, and every point of the other
set lies in the closed region outside $\xi$
(points of $\xi$ are considered both inside and outside).
 In this paper we restrict
our consideration to elements of $\cal C$ being circles.
A circle $C(X,r)$, with center $X$ and radius $r$, separating $P$ from $Q$
is said to be a largest separating circle if there is a
neighborhood $B$ of $X$ such that
there is no 
separating circle with  radius strictly greater
than $r$ centered at a point in $B$.
We propose an optimal algorithm to find all largest circles separating
two given sets of line segments $P$ and $Q$. 

Some previous research on this subject concerned polygonal
separability\cite{ep-mps-88,abosy-fmcnp-89,m-idssp-92}
or its extension to 
higher dimensions, where the construction of 
a polyhedron with a small number of faces,
separating two given polyhedra was
considered.\cite{dj-cmcnp-90,ms-sapo-95,bg-aoscf-95}
Line or hyperplane separability of two given
sets of points may be solved using linear programming.\cite{m-lpltw-84}

The problem of circular separability was first considered in the
context of applications in pattern recognition and image processing,
in particular to recognize digital disks.\cite{ka-dd-84,f-spcrd-86}
Kim and Anderson\cite{ka-dd-84} gave a quadratic algorithm to
determine the circular separability of two finite sets of
points. Bhattacharya\cite{b-cspps-88} computed in $O(n \log n)$ time
the set of centers of all circles that separate two
given point sets. O'Rourke, Kosaraju and Megiddo\cite{okm-ccs-86}
presented optimal algorithms for the circular separability of point
sets.  {They}
determine the circular
separability of two given point sets
and find the smallest  separating circle in linear time
and 
all the largest separating circles in $O(n \log n)$ time. Their method
is based on a well-known transformation that lifts the points on a
paraboloid in 3-space and reduces the  smallest separating circle
problem for two point sets to a convex quadratic minimization problem
in three dimensions.  This method generalizes to spherical
separability in higher dimensions.  However it does not apply to the
problem of circular separability of line segments.
{
The problem of circular separability of two polygons has been 
considered and the smallest separating circle
can be found in linear time.\cite{bcdy-csp-95}
}

In the present paper, we consider the problem of finding all largest circles 
separating two given sets of line segments whose relative interiors
do not intersect.
An $O(n \log n)$ algorithm is given
to solve this problem.
As our algorithm works in the case where
 segments degenerate to single points,
it may be considered as a generalization of the result of
O'Rourke, Kosaraju, and Megiddo\cite{okm-ccs-86}
to line segments.

Our algorithm can be adapted to work in the general case where line
segments may intersect.  In this case, it works in $O(n \alpha(n)
\log^2 n)$ deterministic time or in $O(n \alpha(n) \log n)$ randomized
time and requires $O(n \alpha(n))$ space, where $\alpha(n)$ is the extremely
slowly growing inverse of the Ackermann function.

{
\noindent{\bf Overview of the paper}

The paper is organized as follows.  In section \ref{preliminaries}, we
summarize some results about hierarchical decompositions of convex
polyhedra and the representation of circles in the plane as points of
a three dimensional space called the {\em space of circles}.  Section
\ref{circles} establishes a characterization of separating circles
that are locally maximum.  In section \ref{cones}, we generalize the
hierarchical decompositions presented in section \ref{preliminaries}
and show how to find the intersections between a query line and some
non polyhedral objects in logarithmic time.  All these ingredients are
combined in section \ref{algorithm} to yield an algorithm whose
complexity is shown to be $O(n\log n)$ in section \ref{complexity}.  }

\section{Preliminaries \label{preliminaries}}

{
\noindent{\bf Hierarchical decompositions}
}

We will refer to the {\em hierarchical representation} of convex polygons
introduced  by Dobkin and Kirkpatrick.\cite{k-osps-83,dk-ladsc-85}
Originally, such a representation has been introduced  for planar maps to solve 
the point location problem in optimal
$O(\log\ n)$ time.

Hereafter, we  use outer hierarchical representations. 
An outer  hierarchical representation
of a (possibly unbounded)
 convex polyhedron $D$  is  a nested sequence 
$D_0 \supset D_1 \supset \ldots \supset D_k$ of (possibly unbounded)
 convex polyhedra, such that
\begin{enumerate}
\item [1.] $D_0$ {has constant size (e.g. 4),} 
\item [2.] $D_k$ is the polyhedron D,
\item [3.] the set $F_i$ of faces of $D_i$ is obtained from $F_{i+1}$ 
 by removing a subset $I_{i+1}$ of pairwise nonadjacent faces of $D_{i+1}$. 
 Extending the remaining faces $F_{i+1} \setminus I_{i+1}$
 will then form the polyhedron $D_i$.
\end{enumerate}
It may be proved that, given a convex polyhedron $D_{i+1}$, it is
possible to find a constant fraction of its faces that have a bounded
number of edges and that are pairwise nonadjacent. As a consequence,
the hierarchical representation of a convex polyhedron $D$ with $n$
vertices has  depth $k = O(\log n)$. The whole hierarchical
representation requires $O(n)$ space and can be computed in $O(n)$
time.  After computing the hierarchical representation of a convex
polyhedron, line intersection queries may be performed in $O(\log\ n)$
time.

{
\noindent{\bf Space of circles}
}

The paper will use a well-known transformation $\Phi$, mapping circles
in the $xy$-plane (called {\em horizontal}) to points in the
three-dimensional space which we will call the {\em space of
circles}. According to this transformation, the image of a circle of
radius $r$, centered at $(x_0, y_0)$, is the point $(x_0, y_0, r)$.
Observe that the space of circles is in fact a halfspace, as it
contains only points with non-negative $z$-coordinate. The images of
the circles passing through a point $(x_1, y_1)$ lie on the surface of
a cone of revolution with vertical axis whose apex is the point
$(x_1,y_1,0)$ and whose angle is $45^{\circ}$. Such a cone will be
called a {\em lifting cone} and denoted by $LC(x_1, y_1)$. Notice
that the image of a circle tangent to a given line $l$ lies in a
halfplane containing $l$ and making a $45^{\circ}$ angle with the
$xy$-plane. There are two such {\em lifting halfplanes}, $H^-(l)$ and
$H^+(l)$, denoting the images of the circles tangent to $l$ and
centered, respectively, on the left- or the right-hand side of the
oriented line $l$.

Let $S$ denote the set of line segments $s_1, s_2,\ldots, s_m$ in the
plane.  The closest site Voronoi diagram of $S$, noted $Vor(S)$, is
the partition of the plane into $m$ regions, such that any point
belonging to the $i$-th region is closer to $s_i$ than to any other
segment of $S$. Suppose that we want to decide whether a query disk
contains a point of a given set $S$. Such a query may be answered
quickly if the closest site Voronoi diagram of the set $S$ has been
precomputed. We first locate the Voronoi cell that contains the center
of the query disk, which determines the
closest segment $s_i$ of $S$.  The radius of the disk is then compared
to the distance from its center to $s_i$.

Similarly, in order to decide whether a query disk  entirely contains
a given set of line segments $S$, we will precompute the furthest site
Voronoi diagram of $S$, noted $FVor(S)$,
which is just the furthest site
Voronoi diagram of 
the vertices of the convex hull $CH(S)$ of $S$. 

For the purpose of the paper, it is useful to introduce the following
three-di\-men\-sio\-nal structure, which encapsulates all the
information contained in the furthest site Voronoi diagram $FVor(S)$.
For each vertex $v$ of $CH(S)$, consider the cone $LC(v)$ and let
$UE(S)$ denote the upper envelope of all such cones. A point of
$UE(S)$ corresponds to a circle that encloses $S$ and touches $S$ at
some point.  {Notice that $UE(S)$ is also the boundary of the
intersection of the cones, which is convex.  $UE(S)$ consists of conic
faces glued together along hyperbolic edges.  These arcs are contained
in a vertical plane and projects onto the edges of $FVor(S)$.}

\section{ Largest Separating Circles.\label{circles}}

In the sequel, a segment is said to lie  inside (resp. outside)
a given circle $C$ if it is included in the closed region that is inside
(resp. outside) $C$; such  a  segment and the circle $C$
are allowed to be tangent, i.e. to meet at a single point.
Two sets of segments $P$ and $Q$ are said to be in general
position if they do not admit parallel segments
and if  there is no circle tangent to four segments
of $P \cup Q$.


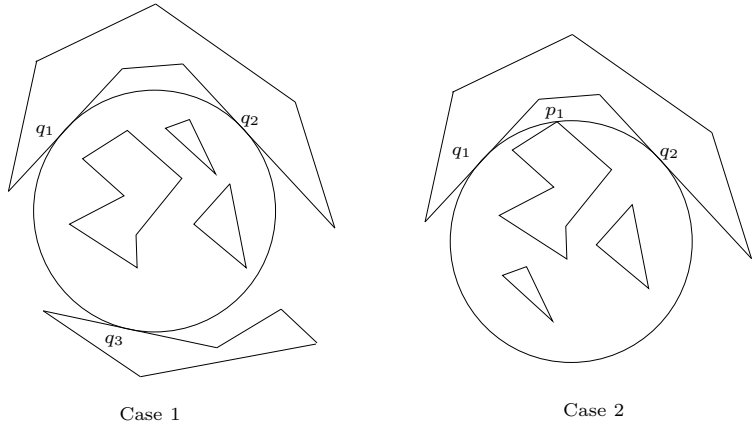
\begin{figure}[tp] \begin{center} 
\ifx\cs\tempdima \newdimen \tempdima\fi
\ifx\cs\tempdimb \newdimen \tempdimb\fi
\ifx\cs\jpdrawx \newcount \jpdrawx\fi
\jpdrawx=100 \tempdima=   1.00cm \tempdimb= 1.0cm
\ifdim \unitlength = 1pt
\unitlength=  1.00cm
\fi
\ifdim \unitlength < 0cm
\multiply \unitlength by -10 \divide \unitlength by 99 \fi
\ifdim \unitlength > \tempdima \tempdima=\unitlength \fi
\ifdim \unitlength < \tempdima \tempdima=\unitlength \fi
\multiply \jpdrawx by \tempdima \divide \jpdrawx by \tempdimb
\setlength{\unitlength}{\tempdima}
\begin{picture}(   9.9,   5.5)( 0.348, 7.555)
\thinlines
\large
\put( 0.348, 7.555){
}
\scriptsize
\put(6.26,11.08){$q_1$}
\put(3.45,11.50){$q_2$}
\put(1.64,8.57){$q_3$}
\put(9.02,11.06){$q_2$}
\put(0.72,11.37){$q_1$}
\put(7.50,11.60){$p_1$}
\put(1.84,7.55){Case 1}
\put(7.74,7.60){Case 2}
\end{picture}
\setlength{\unitlength}{1cm}
\caption{\label{Local}For Lemma \protect\ref{largest-constraints}}
\end{center} 
\end{figure}

\begin{thlemma}
\label{largest-constraints}
If $C$ is a largest circle separating two given sets of segments
 $P$ and $Q$ in general position,
$P$ lying inside $C$ and $Q$ lying outside $C$, then one of the following
two conditions must be verified (see Figure \ref{Local})~:
\begin{enumerate}
\item [$1$.] $C$ is tangent to three segments of $Q$ at  points $q_1$, $q_2$ and $q_3$
 such that all
 three arcs of $C$ determined  by these points are smaller than a semi-circle
 (see Fig. 1, case 1).
\item [$2$.] $C$ is tangent to two segments of $Q$ at points $q_1$ and $q_2$, and meet
  the convex hull $CH(P)$ at a vertex $p_1$,
  such that the arc $q_1q_2$ of $C$ that passes 
through $p_1$  is smaller than a semi-circle, (see Fig 1, case 2).
\end{enumerate}
\end{thlemma}

{
\proof{
Consider a separating circle $C$ with $P$ inside $C$ and $Q$ outside
$C$. We will deform  $C$  until it becomes locally maximal.

We first grow $C$ without moving its center until it touches
$Q$ at a point $q_1$ of some segment $s_1$.

We then grow the circle so that it remains tangent to $s_1$
at $q_1$.  At some stage, the circle hits
$Q$ at a point $q_2$ of a segment $s_2$.

We now keep the circle tangent to  $s_1$ and
$s_2$ and increase its radius.
Either the smaller arc between $q_1$ and
$q_2$ will hit a point of $P$,  in which case Condition 2 holds,
or the circle will hit a segment $s_3$ of $Q$. If
Condition 1 holds, we are done. Otherwise, we exchange the role of $s_1$ or $s_2$ and $s_3$ and continue growing the circle.
} 
}

If $Q$ admits parallel segments, a largest separating circle may be
tangent to two parallel segments of $Q$ without meeting $P$ or
touching a third segment in $Q$ ( see Fig. 2, case $1'$ or $2'$).  In
such a case, there is an infinite number of largest separating circles
that are tangent to those two segments of $Q$.  However all those
circles can be deduced by translation from two extremes circles which, in
addition to the two contact points with parallel segments of $Q$, have
a third contact point with $P$ or $Q$. Our algorithm
 reports only the largest separating circles that have at
least three contact points.  When point sets $P$ and $Q$ are not in
general situation, those circles may be in one of the degenerate cases
listed in the lemma below.

\begin{thlemma}
\label{largest-constraints-bis}
Let $C$ be a largest circle separating  two sets of segments
 $P$ and $Q$, such that
$P$ lies inside $C$, $Q$ lies outside $C$
and $C$ has at least three contact points
with $P$ and $Q$. Then, in addition to cases~1 and 2 of 
Lemma~\ref{largest-constraints} above, $C$ may be in one of the following
degenerate cases (see Fig. 2)~:
\begin{enumerate}
\item [$1'$.]
$C$ is tangent to two parallel segments  of $Q$ 
  (at two diametral 
   points)  and to a third segment in $Q$.
\item [$1''$.] $C$ touches $Q$ at two pairs of
antipodal points.
\item [$2'$.]
$C$ is tangent to two parallel segments  of $Q$ and meet
a vertex of $CH(P)$.
\item [$2''$.] $C$ touches $Q$ at two diametral points $q_1$
and $q_2$ and $P$ at two vertices $p_1$ and $p_2$
such that the points $p_1$, $q_1$, $p_2$, and $q_2$
appear in that cyclic order on circle $C$.
\end{enumerate}
\end{thlemma}

{
\proof{
Easy generalization of proof of Lemma \ref{largest-constraints}.
} 
}

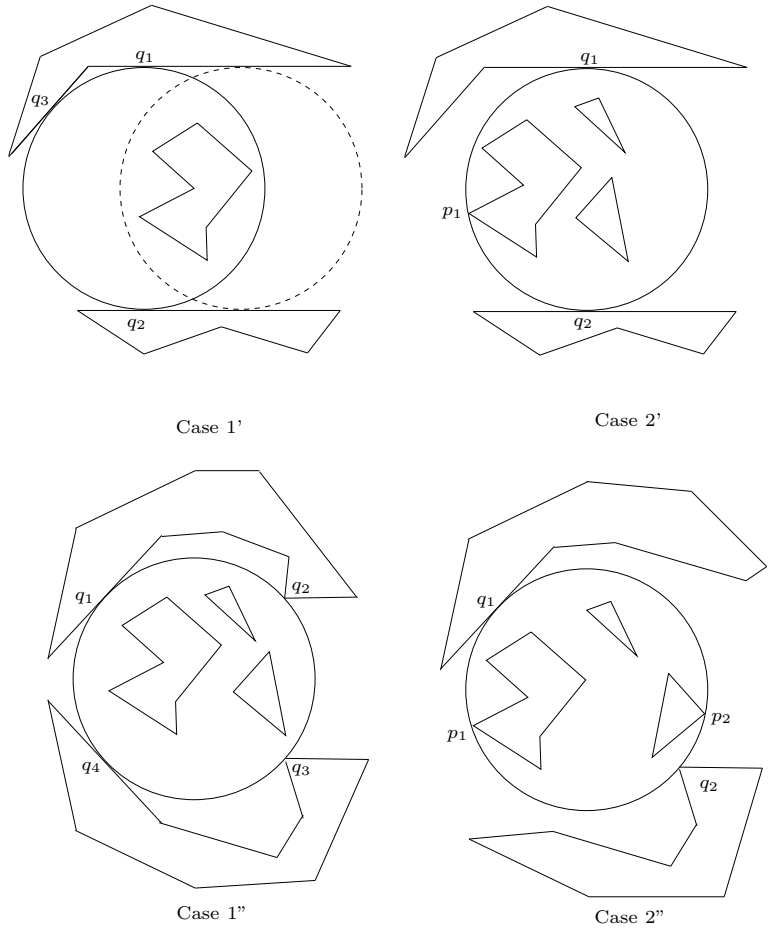
\begin{figure}[tp] \begin{center} 
\ifx\cs\tempdima \newdimen \tempdima\fi
\ifx\cs\tempdimb \newdimen \tempdimb\fi
\ifx\cs\jpdrawx \newcount \jpdrawx\fi
\jpdrawx=100 \tempdima=   1.00cm \tempdimb= 1.0cm
\ifdim \unitlength = 1pt
\unitlength=  1.00cm
\fi
\ifdim \unitlength < 0cm
\multiply \unitlength by -10 \divide \unitlength by 122 \fi
\ifdim \unitlength > \tempdima \tempdima=\unitlength \fi
\ifdim \unitlength < \tempdima \tempdima=\unitlength \fi
\multiply \jpdrawx by \tempdima \divide \jpdrawx by \tempdimb
\setlength{\unitlength}{\tempdima}
\begin{picture}(  10.1,  12.2)( 0.000, 1.145)
\thinlines
\large
\put( 0.000, 1.145){
}
\scriptsize
\put(7.51,9.09){$q_2$}
\put(0.29,12.06){$q_3$}
\put(2.22,7.67){Case 1'}
\put(7.79,7.76){Case 2'}
\put(3.76,5.57){$q_2$}
\put(3.76,3.15){$q_3$}
\put(0.88,5.44){$q_1$}
\put(1.57,9.08){$q_2$}
\put(1.68,12.64){$q_1$}
\put(7.60,12.61){$q_1$}
\put(5.76,10.56){$p_1$}
\put(0.97,3.18){$q_4$}
\put(6.22,5.39){$q_1$}
\put(9.18,2.94){$q_2$}
\put(5.83,3.61){$p_1$}
\put(9.34,3.84){$p_2$}
\put(2.23,1.19){Case 1''}
\put(7.79,1.15){Case 2''}
\end{picture}
\setlength{\unitlength}{1cm}
\caption{\label{Degenerate}Degenerate cases of Lemma \protect\ref{largest-constraints-bis}}
\end{center} \end{figure}

\section{Intersecting Upper Envelope of Cones.\label{cones}}


For any set $S = \{s_1, s_2,\ldots, s_n\}$
 of points in the $xy$-plane  
let $LC(S)$ denote the family of  cones
$LC(s_1), LC(s_2),\ldots,LC(s_n)$ and $UE(S)$ their
upper envelope.
In this section, we will  adapt the hierarchical
representation of polyhedra to obtain the following theorem~:

\begin{theorem}
\label{line-inter}
Let $LC(S)$ be a family of  cones and $F$ be 
{the set of lines of the three-dimensional space.}
It is possible to preprocess  $LC(S)$
in $O(n \log\ n)$ time and $O(n)$ space so that the intersections of $UE(S)$ 
with a query line $l\in F$ can be found in $O(\log\ n)$ time.
\end{theorem}

\proof{ According to an earlier observation, the projection on
the $xy$-plane of the edges of $UE(S)$ is a planar straight-line
subdivision of size $O(n)$ whose cells are unbounded.  In linear time,
it is possible to find a fraction of its faces that are pairwise
nonadjacent and such that each face has a bounded number of edges.
After eliminating from $S$ the apexes of the cones corresponding to
those faces we are left with a subset $S'$ of $S$.  By repeating this
process, we construct a hierarchical representation $S_1
\subset S_2 \subset\ldots\subset S_k=S$, where $S_1$ is a single
point. We obtain as well a hierarchy $G(S_1)$, $G(S_2)\ldots G(S_k)$
of $k = O(\log\ n)$ straight-line planar subdivisions, such that any
face of $G(S_i)$ intersects a bounded number of faces of $G(S_{i-1})$ and
vice versa. Using Kirkpatrick's technique,\cite{k-osps-83} this
hierarchical representation can be found in $O(n)$ time and requires
$O(n)$ space.

{As already observed, $UE(S)$ is the boundary of
the intersection of the cones $LC(S)$,
which is  convex. As a consequence, a line}
$l$ intersects $UE(S)$ in at most two points. In fact, if $l$
intersects $UE(S)$ in two points, it intersects each $UE(S_i)$,
$i=1,2,\ldots,k$, in two points, and, if $l$ intersects $UE(S)$ in a
single point, it intersects each $UE(S_i)$ in a single point. Suppose
that we know an intersection $x$ of a query line $l$ with
$UE(S_{i-1})$ and the face of $G(S_{i-1})$ that contains the
projection $x'$ of $x$.  In constant time, we can
compute the corresponding intersection of $l$ with $UE(S_i)$, as well
as the face of $G(S_i)$ that contains its projection. In $k = O(\log\ n)$
steps, each one taking a constant time, we can compute the
intersection of $l$ with $UE(S)$.  } 

Note that, as $UE(S_i)$ is contained in $UE(S_{i-1})$, it may happen that $l$ 
intersects $UE(S_{i-1})$ but not $UE(S_i)$.

Suppose that, instead of a line, the query curve 
$\zeta \in F$ belongs to one of the three following categories:
\begin{enumerate} 
  \item  $\zeta = LC(x_1, y_1) \cap LC(x_2, y_2)$ is the intersection of two 
      cones for  two points $(x_1, y_1)$ and $(x_2, y_2)$ of the 
     $xy$-plane,
  \item $\zeta = LC(x_1, y_1) \cap H^+(l)$ is the intersection of a  cone 
     and a  halfplane, for a point $(x_1, y_1)$ and an oriented 
     line $l$ of the $xy$-plane,
  \item  $\zeta = H^+(l_1) \cap H^+(l_2)$ is the intersection of  two  
     halfplanes.   
\end{enumerate}
Family $F$ contains now branches of hyperbola, parabolas and lines.
The above theorem generalizes then to 

\begin{theorem}
\label{curve-inter}
Let $F$ be a family of curves in three dimensions such that any curve 
$\zeta \in F$ is an intersection of two surfaces, each of these surfaces
being either a  cone or a  halfplane. 
\begin{enumerate}
  \item There are at most two intersections of $\zeta \in F$ with
    $UE(S')$, for any $S' \subseteq S$.
  \item It is possible to preprocess the cones $LC(S)$ in $O(n)$ 
    time, using $O(n)$ space, so that the intersections of $UE(S)$ with 
    a query curve $\zeta \in F$ can be found in $O(\log n)$ time. 
\end{enumerate}
\end{theorem}

{
\proof{
Let $\zeta \in F$.  $\zeta$ is the image by $\Phi$ of the set of
circles tangent to $t$ and $t'$ where $t$ and $t'$ are either points
or lines.  Let $q$ be a moving point on $\zeta$ and $p\in S$.  Imagine
that $q$ moves along $\zeta$ starting from infinity. At some position
$q=q_p^{in}$, $p$ enters the circle $\Phi^{-1}(q)$ and, at some
position $q=q_p^{out}$, $p$ gets out of the circle and remains outside
the circle while $q$ goes towards the other point at infinity on
$\zeta$.  Thus $\zeta$ intersects $LC(p)$ in two points $q_p^{in}$ and
$q_p^{out}$.  Now, consider the intersection of $\zeta$ with $UE(S')$.
Consider as above a moving point $q$ on $\zeta$. Let $q_{S'}^{in}$ be
the last of the $q_p^{in}$ and let $q_{S'}^{out}$ be the first of the
$q_p^{out}$.  At $q_{S'}^{in}$, $q$ has entered all the cones $LC(p)$
for all $p\in S'$ and, at $q_{S'}^{out}$, $q$ gets out of one of the
cones $LC(p)$ for some $p\in S'$ and will never get in again.  Thus if
$q_{S'}^{in}$ lies before $q_{S'}^{out}$ along $\zeta$, then $\zeta$
intersects $UE(S')$ twice, otherwise $\zeta\cap UE(S') =\emptyset$.
Point {\it 1} of the lemma is proved.

Point {\it 2} is a direct generalization of Theorem \ref{line-inter}.
Given the hierarchical decomposition of $S=S_k\supset\ldots S_2\supset S_1$,
if the intersection between $\zeta$ and $UE(S_{i-1})$ 
is known, the possible intersections
between $\zeta$ and $UE(S_i)$ can clearly be determined in constant time.
} 
}

\section{The Algorithm.\label{algorithm}}

Before turning our attention to the algorithm, we make a few simple
observations about the images 
of some families of circles in the space of circles.

\begin{fact}
\label{hyperbola}
The image (by $\Phi$) of a family of circles 
passing through two given points $s_1$ and $s_2$ is the branch of
hyperbola $LC(s_1) \bigcap LC(s_2)$.
\end{fact}

\begin{fact}
\label{parabola}
The image (by $\Phi$) of a family of circles 
tangent  to a given oriented line 
$l_1$ lying on the right of $l_1$,
 and passing through a given point $s_1$
 is the parabola $LC(s_1) \bigcap H^+(l_1)$.
\end{fact}

\begin{fact}
\label{line}
The image (by $\Phi$) of a family of circles 
tangent to two given oriented lines
$l_1$ and $l_2$ and  lying on the right of $l_1$ and $l_2$
is the line $H^+(l_1)\bigcap H^+(l_2)$.
\end{fact}

To find the largest circles separating two sets of line segments $P$
and $Q$, we will run the algorithm given below twice.  The algorithm
looks first for the largest separating circles $C$ enclosing $P$
and in a second
run, the roles of $P$ and $Q$ are exchanged. The algorithm
will report all largest separating circles with at least
three contact points.

The idea of the algorithm is to search  all the circles that
 verify one of the conditions of Lemmas \ref{largest-constraints}
or \ref{largest-constraints-bis}. 
Consider first Conditions $1$, $1'$ or $1''$.
Any circle $C$ tangent to
$Q$ at three points and not containing any point of $Q$ in its interior is
centered at a vertex of $Vor(Q)$, the closest site Voronoi diagram of the
set of line segments $Q$. For each such vertex $v$,
we determine which face of
$FVor(P)$ it belongs to.
This way, we can compute the distance from $v$ to 
its furthest point in $P$. 
If this distance appears to be smaller that the radius
of the Voronoi circle $C$ centered at $v$,
 $C$ separates $P$ and $Q$. In such a case, if $C$ verifies conditions
$1$, $1'$ or $1''$,
it is reported as a  largest separating circle. 

When the separating circle $C$ verifies the condition $2$ of Lemma
\ref{largest-constraints} or one of the degenerated conditions $2'$
and $2''$ of Lemma \ref{largest-constraints-bis}, it must be tangent
to $CH(P)$ at some vertex $p_1$, and tangent to $Q$ at two points
$q_1$ and $q_2$.  The first condition means that $\Phi(C)$ lies on
$UE(P)$, within the face corresponding to vertex $p_1$.  At the same
time, the center of $C$ lies on a Voronoi edge of $Vor(Q)$
equidistant from $q_1$ and $q_2$.  Suppose that $q_1$ and $q_2$ are
internal points of two edges of $Q$, then, in the space of circles,
$\Phi(C)$ lies on a segment whose supporting line is determined according
to Fact \ref{line}.  Similarly, if $q_1$ or $q_2$ are endpoints of
segments of $Q$, the corresponding edge of $Vor(Q)$ is mapped in the
space of circles to a parabola segment or to a hyperbola segment as stated
in Facts \ref{hyperbola} and
\ref{parabola}. Thus, to find the largest separating circles
that fulfill conditions $2$, $2'$ or $2''$, it is sufficient to
examine in turn all the $O(n)$ edges of $Vor(Q)$. For each edge of
$Vor(Q)$, we compute the intersection of the line, parabola or
hyperbola segment that is the image of the largest circles centered on
this edge with the envelope $UE(P)$.  The hierarchical representation
of $UE(P)$ is used for this purpose. Each point of intersection
corresponding to a circle satisfying one of the conditions $2$, $2'$
or $2''$ is reported as a largest separating circle.

\vbox{
\begin{algorithm}{All Largest Separating Circles}\\
\begin{enumerate}
  \item [{\bf Input:}]  Two sets of line segments $P$ and $Q$ 
    with  a total 
     of $n$ segments whose relative interiors do not intersect.\\
  \item [{\bf Output:}] All largest separating circles $C$, with
$P$ inside $C$, $Q$ outside $C$, and at least three contact points.
\end{enumerate}
\begin{enumerate}
  \item Compute $FVor(P)$, the furthest site Voronoi diagram of the vertices
     of the convex hull $CH(P)$
          of set $P$; compute $UE(P)$, the image of $FVor(P)$ in the 
     space of circles.

  \item Compute the hierarchical representation of $UE(P)$.

  \item Compute $Vor(Q)$, the closest site Voronoi diagram of the set $Q$.

  \item {\bf for} each vertex $v$ of $Vor(Q)$
    \begin{itemize}
       \item [4.1.] Compute the distance $d(v,Q)$ from $v$ to $Q$.
       \item [4.2.] Locate $v$ in a face of $FVor(P)$ and compute 
        $d(v,FVor(P))$, the
        distance from $v$ to its most distant vertex in $P$.
       \item [4.3.] {\bf if} $d(v, Q) \geq d(v, FVor(P))$ {\bf and}
                     one of the conditions $1$, $1'$ or $1''$ holds for the
		   circle $C$ centered at $v$ with radius $d(v, Q)$
           {\bf then Output($C$)}.
    \end{itemize}
  \item {\bf for} each edge $e$ of $Vor(Q)$
    \begin{itemize} 
       \item  [5.1.] Compute the curve segment $z$ in the space of circles 
             that is the image of the  two largest circles centered on $e$ and
              tangent to $Q$. Let  $\zeta$  be the curve (line, parabola
			 or hyperbola) supporting $z$.
       \item  [5.2.] Compute $x_1$ and $x_2$, the at most two
	    				intersections of $\zeta$ with $UE(CH(P))$ if they exist.
       \item  [5.3.] {\bf for} $i=1,2$
					 {\bf if} $x_i\in z$ {\bf and}  $x_i$ is the image of a
					  circle $C_i$ such that conditions $2$, $2'$ or $2''$  hold
					 {\bf then Output($C_i$)}.
    \end{itemize}
  \end{enumerate}
\end{algorithm}
}

The correctness of the algorithm directly follows from Lemmas
\ref{largest-constraints}, \ref{largest-constraints-bis} and the previous
discussion.

\section{Complexity of the Algorithm.\label{complexity}}

The computation of the furthest site Voronoi diagram in step 1 takes 
$O(n \log n)$ time and $O(n)$ space by well known algorithms.\cite{k-eccs-79}
The upper envelope $UE(P)$ is obtained in $O(n)$ time by
lifting each face and edge of $FVor(P)$ onto the corresponding face
and edge of $UE(P)$.

The hierarchical representation of $UE(P)$ in step 2 is computed in 
$O(n)$ time using $O(n)$ space by Theorem \ref{curve-inter}.

The Voronoi diagram of the set of line segments in step 3 can be computed in
$O(n \log\ n)$ time  using $O(n)$ space.\cite{ld-gvdp-81}

The for loop in step 4 is run $O(n)$ times. Step 4.1 takes a constant
time. The hierarchical representation of $FVor(P)$ computed in step 2
can be used to perform the point locations of step 4.2 in $O(\log n)$
time per query. \cite{k-osps-83} Step 4.3 requires time proportional
to the degree of vertex $v$ and the time complexity of step 4.3,
over all iterations of the for loop, is $O(n)$. The total
time complexity of the for loop in step 4 is $O(n \log n)$.

Similarly, the loop in step 5 is executed $O(n$) times. Depending
on the case, the curve segment $z$ needed in step 5.1 is computed
using one of the Facts \ref{hyperbola}, \ref{parabola} or \ref{line}.
By Theorem \ref{curve-inter}, there are at most two intersections of
$z$ with $UE(CH(P))$ and they can be computed in $O(\log n)$ time.
Step 5.3 requires constant time. We conclude that step 5 takes
$O(n \log\ n)$ time.

We have thus proved

\begin{theorem}
\label{algo-compl}
Given two sets of line segments $P$ and $Q$ with a total of $n$
segments whose relative interiors do not intersect, it is possible to
compute all largest circles separating $P$ and $Q$ in $O(n\ \log\ n)$
time using $O(n)$ space.
\end{theorem}

Once those largest separating circles have been found, the largest one
can easily be reported by comparing the radii.

\begin{figure}[tp] \begin{center} {\def\IPEfile{Lower.ipe}\begingroup
  \catcode`\%=9\catcode`\!=0\catcode`\-=11\input{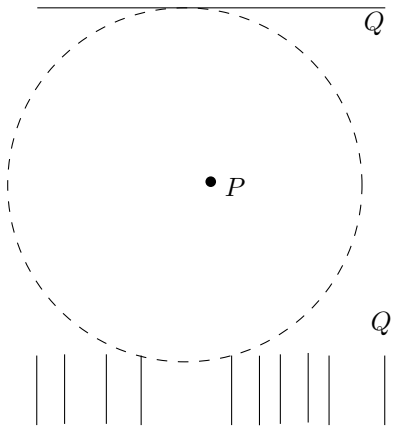}}
\caption{\label{Lower}The lower bound example}
\end{center} \end{figure}

In order to show that our result is optimal, we sketch the proof of a
$\Omega (n\ \log\ n)$ lower bound for our problem.\cite{okm-ccs-86}
We proceed by reduction
to the maximum gap problem for which $\Omega (n\ \log\ n)$ is a lower bound
in the linear decision-tree model of
computation.\cite{mt-cppna-85} Let $X = \{x_1, x_2,
\ldots, x_n\}$ be a set of points on the real line between
$x_{min}$ and $x_{max}$ for which the maximum gap must be computed,
i.e. the largest interval between two consecutive points of X. Let set
$Q$ contain $n$ line segments, each one extending between the points
$(x_i, -1)$ and $(x_i, 0)$, $i=1, 2, \ldots ,n$ and the $(n+1)$-th
segment $s$ extending between the points $(x_{min}, x_{max}-x_{min})$
and $(x_{max}, x_{max}-x_{min})$.  Let $P$ consist of a single point of
coordinates $\left( \frac{x_{min}+x_{max}}{2} ,
\frac{x_{max}-x_{min}}{2} \right)$.  Clearly, the largest circle
separating $P$ and $Q$ is tangent to $s$ and passes through segments
at $x_i$ and $x_j$ defining the maximum gap in $X$ (see Figure
\ref{Lower}).  { In this construction, the set of segment is not in
general position, but if we symbolically perturb the segments, we will
find one of the maximal gaps.}

It follows from the algorithm that there are at most $O(n)$ largest
separating circles with at least three contact points.  Indeed, for
each of the $O(n)$ vertices of $Vor(Q)$ there is at most one such
circle, and for each of the $O(n)$ edges of $Vor(Q)$ there are at most
two such circles. The above example where the $x_i$ are equally spaced
shows that there are sets $P$ and $Q$ that actually admit $O(n)$
largest separating circles.

\newpage

\section{Conclusions.}

The paper gives an efficient algorithm for the problem of finding all
largest circles separating two given sets of line segments.  The
solution is optimal in the linear decision-tree model of computation.
However, our result does not imply a $\Omega (n\ \log\
n)$ lower bound for the problem of computing the largest circle separating two
given polygons since it is not possible to build a
polygon from a set of line segments  in linear time.

It was supposed in this paper that the relative interiors of the line
segments do not intersect. For two arbitrary sets of segments, we may
have $\Omega(n^2)$ points of intersection.  However, the following
corollary states that we can tackle the problem of determining a
largest separating circle in
less than quadratic time.

\begin{thcorollary}
For two sets of line segments $P$ and $Q$ containing a total of $n$
line segments, it is possible to compute all locally largest circles
separating $P$ and $Q$ in $O(n \alpha(n) \log^2 n)$ deterministic time
or in $O(n \alpha(n) \log n)$ randomized time using $O(n \alpha(n))$
space.
\end{thcorollary}

To prove this, observe that if there exists a circle $C$ separating
two sets of line segments $P$ and $Q$, with $P$ inside $C$ and $Q$
outside $C$, $C$ separates the boundary of the unbounded cell of the
arrangement of the line segments of $P$ from the cell of the
arrangement of the line segments of $Q$ that contains the vertices of
$P$.  The complexity of such a cell is at most $O(n
\alpha(n))$\cite{egs-cmcap-90} and it can be computed in $O(n
\alpha(n) \log^2 n)$ deterministic time\cite{cegss-cfals-93} or in
$O(n \alpha(n) \log n)$ randomized time.\cite{bds-lric-95} Once both
cells have been computed, we can apply our algorithm to $O(n
\alpha(n))$ portions of line segments whose relative interiors do not
intersect.

An interesting open problem is to extend the algorithm to  other 
classes of objects like, for example, circles or figures bounded by line 
segments and circular arcs.



\section{Bibliography}

\begin{thebibliography}{10}

\bibitem{abosy-fmcnp-89}
A.~Aggarwal, H.~Booth, J.~O'Rourke, Subhash Suri, and C.~K. Yap.
\newblock Finding minimal convex nested polygons.
\newblock {\em Inform. Comput.}, 83(1):98--110, October 1989.

\bibitem{b-cspps-88}
B.~K. Bhattacharya.
\newblock Circular separability of planar point sets.
\newblock In G.~T. Toussaint, editor, {\em Computational Morphology}, pages
  25--39. North-Holland, Amsterdam, Netherlands, 1988.

\bibitem{bcdy-csp-95}
J.-D. Boissonnat, J. Czyzowicz, O. Devillers, and M.
  Yvinec.
\newblock Circular separability of polygons.
\newblock In {\em Proc. 6th ACM-SIAM Sympos. Discrete Algorithms}, pages
  273--281, 1995.

\bibitem{bg-aoscf-95}
H.~Br{\"o}nnimann and M.~T. Goodrich.
\newblock Almost optimal set covers in finite {VC}-dimension.
\newblock {\em Discrete Comput. Geom.}, 14:263--279, 1995.

\bibitem{cegss-cfals-93}
B. Chazelle, H.~Edelsbrunner, L.~J. Guibas, M. Sharir, and
  J.~Snoeyink.
\newblock Computing a face in an arrangement of line segments and related
  problems.
\newblock {\em SIAM J. Comput.}, 22:1286--1302, 1993.

\bibitem{dj-cmcnp-90}
G.~Das and D.~Joseph.
\newblock The complexity of minimum convex nested polyhedra.
\newblock In {\em Proc. 2nd Canad. Conf. Comput. Geom.}, pages 296--301, 1990.

\bibitem{bds-lric-95}
M.~de~Berg, K.~Dobrindt, and O.~Schwarzkopf.
\newblock On lazy randomized incremental construction.
\newblock {\em Discrete Comput. Geom.}, 14:261--286, 1995.

\bibitem{dk-ladsc-85}
D.~P. Dobkin and D.~G. Kirkpatrick.
\newblock A linear algorithm for determining the separation of convex
  polyhedra.
\newblock {\em J. Algorithms}, 6:381--392, 1985.

\bibitem{egs-cmcap-90}
H.~Edelsbrunner, L.~J. Guibas, and M. Sharir.
\newblock The complexity of many cells in arrangements of planes and related
  problems.
\newblock {\em Discrete Comput. Geom.}, 5:197--216, 1990.

\bibitem{ep-mps-88}
H.~Edelsbrunner and F.~P. Preparata.
\newblock Minimum polygonal separation.
\newblock {\em Inform. Comput.}, 77:218--232, 1988.

\bibitem{f-spcrd-86}
S.~Fisk.
\newblock Separating points by circles and the recognition of digital discs.
\newblock {\em IEEE Trans. Pattern Anal. Mach. Intell.}, 8(4):554--556, 1986.

\bibitem{ka-dd-84}
C.~E. Kim and T.~Anderson.
\newblock Digital disks.
\newblock {\em IEEE Trans. Pattern Anal. Mach. Intell.}, PAMI-6(5):639--645,
  1984.

\bibitem{k-eccs-79}
D.~G. Kirkpatrick.
\newblock Efficient computation of continuous skeletons.
\newblock In {\em Proc. 20th Annu. IEEE Sympos. Found. Comput. Sci.}, pages
  18--27, 1979.

\bibitem{k-osps-83}
D.~G. Kirkpatrick.
\newblock Optimal search in planar subdivisions.
\newblock {\em SIAM J. Comput.}, 12(1):28--35, 1983.

\bibitem{ld-gvdp-81}
D.~T. Lee and R.~L. {Drysdale, III}.
\newblock Generalization of {Voronoi} diagrams in the plane.
\newblock {\em SIAM J. Comput.}, 10:73--87, 1981.

\bibitem{mt-cppna-85}
U.~Manber and M.~Tompa.
\newblock The complexity of problems on probabilistic, nondeterministic, and
  alternating decision trees.
\newblock {\em J. ACM}, 32(3):720--732, 1985.

\bibitem{m-lpltw-84}
N.~Megiddo.
\newblock Linear programming in linear time when the dimension is fixed.
\newblock {\em J. ACM}, 31:114--127, 1984.

\bibitem{ms-sapo-95}
J. S.~B. Mitchell and S. Suri.
\newblock Separation and approximation of polyhedral objects.
\newblock {\em Comput. Geom. Theory Appl.}, 5:95--114, 1995.

\bibitem{m-idssp-92}
D.~M. Mount.
\newblock Intersection detection and separators for simple polygons.
\newblock In {\em Proc. 8th Annu. ACM Sympos. Comput. Geom.}, pages 303--311,
  1992.

\bibitem{okm-ccs-86}
J.~O'Rourke, S.~R. Kosaraju, and N.~Megiddo.
\newblock Computing circular separability.
\newblock {\em Discrete Comput. Geom.}, 1:105--113, 1986.

\end{thebibliography}
\end{document}